\documentstyle[aps,preprint]{revtex}

\begin{document}
\draft

\title{
 LATTICE EFFECTS IN CRYSTAL EVAPORATION
 }

\author{ I.  Pagonabarraga$^{*}$,  J.  Villain, I.  Elkinani  and M.B.
Gordon}
\address{  DRFMC/SPSMS/MDN,  Centre d'Etudes Nucl\'{e}aires de
Grenoble, \\
85X, F-38041  Grenoble  Cedex,  France\\
$^{*}$  Current
address:  Dep.  F\'{\i}sica  Fonamental,  Universitat de  Barcelona,\\
Av.  Diagonal 647, 08028 Barcelona, Spain }
\date{\today}

\maketitle
\begin{abstract}

	We study the  dynamics  of a stepped  crystal  surface  during
evaporation,  using the classical  model of Burton, Cabrera and Frank,
in which the  dynamics  of the surface is  represented  as a motion of
parallel,   monoatomic   steps.  The   validity   of   the   continuum
approximation   treated   by  Frank  is  checked   against   numerical
calculations   and  simple,   qualitative   arguments.  The  continuum
approximation  is  found  to  suffer  from  limitations   related,  in
particular, to the existence of angular points.  These limitations are
often related to an adatom  detachment rate of adatoms which is higher
on the  lower  side of each step  than on the upper  side  ("Schwoebel
effect").

\end{abstract}

\section{Introduction}

	  The  study  of  the  dynamics  of a  crystal  surface  under
nonequilibrium conditions is an old subject, the interest of which was
renewed in the last years due to modern experimental  techniques, such
as electron reflection microscopy \cite{Met} \cite{Lat}, which allow a
more detailed  observation  of dynamic  phenomena of the surfaces both
during  evaporation and growth.  Moreover,  equilibrium is practically
never  reached, and even a surface which is apparently at  equilibrium
on a certain lengthscale, will present nonequilibrium shapes at larger
lengthscales.  Therefore,  it is of interest to  understand  how these
features can be formed, either during growth or during annealing.

	Burton,  Cabrera and Frank  \cite{BCF}  (BCF) have developed a
simple  theory to describe  the growth of a stepped,  dislocation-free
crystal  surface.  They wrote  equations of motion for the steps.  The
purpose of the present article is to solve these equations for special
cases corresponding to a given initial profile of the surface.

	When trying to solve the BCF equations, a possible approach is
to make the continuum approximation.  Then, as will be seen in Section
2, it is  possible to use a theorem  discovered  by Frank  \cite{Fra}.
However, the continuum  approximation  is questionable if the slope of
the surface is  discontinuous.  For that  reason, the present  work is
focussed  on the  evaporation  of  surfaces  which  initially  contain
corners (Fig.\ \ref{fig1} a).

	Only  evaporation  will be considered  here, in order to avoid
problems due to  nucleation  of new terraces  during  growth.  One can
also  notice  that in situ study of surface  dynamics  is more  easily
performed under  evaporation than under growth  conditions  \cite{Met}
\cite{Lat}.  In this paper we investigate,  within the BCF theory, the
evaporation  of a  defect-free  surface  made  of  parallel,  but  not
equidistant  steps.  Our main  motivation  is to test,  in  particular
instances,  the validity of the continuum  approximation  developed by
Frank \cite{Fra}.  The continuum  approximation  will be shown to fail
when curvature  changes abruptly at some place on the surface.  In the
original BCF paper,  atoms  detaching  from a step were  assumed to be
emitted  with  equal  probability  from the upper  and from the  lower
terrace.  This will be called the {\it{symmetric  case}}.  However, as
noticed by Ehrlich  \cite{Ehr},  a step may be  partially  rather than
totally  absorbing.  Generally,  one side is more  absorbing  than the
other \cite{Sch}.  This is known as Schwoebel effect.

	We will consider solid  surfaces made of parallel  steps.  The
structure  of the  surface at time $t$ is fully  characterized  by the
positions  $x_{n}(t)$ of the  successive  steps  labelled by the index
$n=1,2,...$   as  shown  in  Fig.  \  \ref{fig1}   a.  The   functions
$x_{n}(t)$ satisfy the following  equation, which follows from the BCF
theory,  appropriately  modified to take into  account  the  Schwoebel
effect:

\begin{equation}
\dot{x}_{n}=-\varphi_{l}(x_{n}-x_{n-1})-\varphi_{r}(x_{n+1}-x_{n})
							  \label{1}
\end{equation}

	The expressions of $\varphi_{l}(l)$ and  $\varphi_{r}(l)$  are
derived  in  Reference  \cite{B&Z}  and,  for  completeness,   briefly
rederived in Appendix A.  The first one,

\begin{equation}
\varphi_{l}(l)=(\rho_{0}-F\tau_{0}) \frac{\cosh(\kappa l) -1
+\frac{\kappa D}{D'} \sinh(\kappa l)}{\left( 1+\frac{\kappa^{2}
D^{2}}{D'D''} \right) \sinh(\kappa l)+\left(\frac{\kappa
D}{D'}+\frac{\kappa D}{D''}\right) \cosh(\kappa l)}\kappa D
						\label{1var1}
\end{equation}

\noindent  gives the net flux of outgoing  adatoms  from a step to the
upper terrace, and

\begin{equation}
\varphi_{r}(l)=(\rho_{0}-F\tau_{0}) \frac{\cosh(\kappa l) -1
+\frac{\kappa D}{D''} \sinh(\kappa l)}{\left( 1+\frac{\kappa^{2}
D^{2}}{D'D''} \right) \sinh(\kappa l)+\left(\frac{\kappa
D}{D'}+\frac{\kappa D}{D''}\right) \cosh(\kappa l)}\kappa D
						\label{1var2}
\end{equation}

\noindent is the net flux of outgoing adatoms from a step to the lower
terrace.  In these expressions,  $\rho_{0}$ is the equilibrium density
of adatoms on the high symmetry surface at the appropriate temperature
and vapour pressure, $D$ is the surface diffusion  constant of adatoms
and $1/\tau_{0}$ is the evaporation  probability of an adatom per unit
time.  $F$ is zero in the cases  addressed  here,  but it would be the
beam intensity in the case of growth by molecular  beam epitaxy.  $D'$
is the unit time  probability  of an adatom to stick to a step when it
is just aside the step, and $D''$ is the unit time  probability  of an
adatom  to stick to a step  when it is just  above  that  step  (Fig.\
\ref{fig1}  b).  The  situation  where $D'\neq D$ and $D''\neq D$ will
correspond to situations in which the sticking  probability of adatoms
to steps will  depend on the  terrace  towards  which they move.  More
specifically,  we will call  normal  Schwoebel  effect when  particles
mainly move towards the upper terrace,  $D''<<D$,  while the situation
in which  particles  move  towards  the  lower  terrace  will be named
inverse Schwoebel effect.  Finally,

\begin{equation}
\kappa=\frac{1}{\sqrt{D \tau_{0}}}
						 \label{3}
\end{equation}

\noindent is the reciprocal of the average distance on which an adatom
would  diffuse  on  an  infinite  terrace  before   evaporating.  This
distance is always much larger than the interatomic  spacing, taken to
be the  length  unit  throughout  this work,  while the time an adatom
needs to  diffuse  along the  interatomic  distance  is the unit time.
Normal  Schwoebel  effect  is known to  produce  instabilities  during
evaporation  \cite{Sto},  but the present study is restricted to cases
where no such instability occurs.

	As seen from (\ref{1var1}) and (\ref{1var2}), $\varphi_{l}(l)$
and  $\varphi_{r}(l)$  are  proportional  to $l$  for  small  $l$  and
constant for large $l$.  In the symmetric case one has $D'=D''=D$ and,
since $\kappa << 1$, (\ref{1var1}) and (\ref{1var2}) reduce to the BCF
formula

\begin{equation}
\varphi_{l}(l)=\varphi_{r}(l)=\varphi(l)=D\kappa (F\tau_{0}-\rho_{0})
\tanh\left(\frac{\kappa l}{2}\right)
							\label{2}
\end{equation}

	The  conditions  of  applicability  of the BCF  theory,  which
implies the stability of the step flow regime, are the following.

\begin{enumerate}

\item No dislocations  should be present.  Generally this implies that
only a small part of the surface is observed,  e.g.  0.01 cm x 0.01 cm
for Si wafers \cite{Met}.

\item No surface vacancies should be formed on terraces.  This implies
that the temperature should not exceed a certain threshold, e.g.  1200
K for Si(001) \cite{Tro} though much higher for Si(111) \cite{Met1}.

 \item The vicinal  orientation should be stable with respect to facet
formation.  \item The steps should not undergo  instabilities  such as
the one  discovered  by Bales and Zangwill  \cite{B&Z}  in the case of
growth.

\end{enumerate}

       In the next section, the continuum  approximation  of (\ref{1})
will be introduced and exploited.  We will then focus our attention on
the evaporation of a surface  limited  initially by two  semi-infinite
half planes as sketched in Fig.\ \ref{fig1} a.  One of them is assumed
to be a high  symmetry  surface and the other one is made of  parallel
steps.  In  section  3, the long time  behaviour  of the  solution  of
(\ref{1}) will be derived analytically.  In section 4, exact solutions
will be derived for approximations of (\ref{1}).  Numerical  solutions
of (\ref{1})  will be presented in section 5 and will be compared with
the  continuum   approximation.  Finally,  more  complicated   initial
profiles will be investigated in Section 6, namely a periodic  surface
with grooves.  In the  discussion  we summarize  our main results, and
technical details are worked out in the appendices.

\section{The continuum model and Frank construction}

	In the continuum  approximation,  the step position $x_{n}$ is
regarded as a continuous and generally  differentiable function of the
local surface  height $-n$, also  considered as a continuous  variable
$z$.  Therefore,  $x_{n}\rightarrow x(-z)$.  Moreover, introducing the
derivatives $x'=-\partial x/\partial z$, $x''=\partial^{2}  x/\partial
z^{2}$, etc., and  $\varphi'(x')=\partial  \varphi (x')/ \partial x'$,
and assuming  $x'''$ and the further  derivatives to be small, one can
write (\ref{1}) as

\begin{equation}
\dot{x}=-2 \phi (x') +\frac{x''}{2} \psi' (x') - \frac{x'''}{3}
\phi' (x')- \frac{x''^{2}}{4 } \phi'' (x')                 \label{4}
\end{equation}

\noindent where

\begin{eqnarray}
 \phi(x')&=&\frac{\varphi_{l}(-x') +\varphi_{r}(-x')}{2}
 						    \label{5}   \\
\psi(x')&=&\varphi_{l}(-x')-\varphi_{r}(-x')
						\label{psi}
\end{eqnarray}

\noindent  and  $\varphi '$ and $\varphi  ''$ are the  derivatives  of
$\varphi$.

	In this  section, we consider  the  approximation  obtained by
keeping only the first term in (\ref{4}),

\begin{equation}
\dot{x}=-2 \phi (x')                                       \label{4a}
\end{equation}

\noindent   This   approximation   will  be   called   the   continuum
approximation.  It  clearly   fails  near  a  corner,  when  $x''$  is
infinite.  Moreover,  it may be expected to be better when there is no
Schwoebel effect, since then the second term of the right hand side of
eq.(\ref{4}) vanishes according to (\ref{psi}).

	Equation (\ref{4a}) may alternatively be written as a relation
between $z'=( \delta z/\delta x)_{t}=$ $1/(\partial x/\partial z)_{t}$
$=-x'$ and $\dot{z}= (\partial x/\partial z)_{x}=\dot{x}/x'$:

\begin{equation}
\dot{z}=2 z'\phi(\frac{1}{z'})                              \label{6}
\end{equation}

	From this  expression  one can  deduce  the decay  rate of the
surface,  $v$,  at  a  point   characterized  by  a  direction  vector
$\vec{n}=(n_{x},0,n_{z})$,  with  $n_{x}=-1/\sqrt{1+x'^{2}}$.  Due  to
the    geometry,    the    decay    rate   is   given   by    $v\equiv
\dot{z}\;n_{z}=\dot{x}\;n_{x}$, so that

\begin{equation}
v=-2
\frac{\phi(x')}{\sqrt{1+x'^{2}}} \equiv f(\vec{n})          \label{7}
\end{equation}

	This equation shows that in the continuum limit the decay rate
of the surface depends only on its local  orientation.  The problem of
the decay or growth of a surface  when its  velocity  is a function of
the  orientation  has been  investigated  by Frank  \cite{Fra}.  Frank
proved that the points at the surface with a given normal  orientation
move on straight  lines.  Let us consider the initial profile of Fig.\
\ref{fig1} a.  For a velocity given by  eq.(\ref{7}),  we can draw the
polar plot  $\Gamma$ of the  inverse of the  velocity,  as is shown in
Fig.\ \ref{fig2}.  From the curve ($\Gamma$) we can deduce the further
evolution of the profile.  If we take a point $P$ of the crystal where
the  normal is  $\vec{n}$,  it  determines  a point $M$ of  ($\Gamma$)
through  $\vec{OM}//\vec{n}$.  Then  Frank's  theorem  states that the
point $P$ moves on a straight line  parallel to the normal  $\vec{n}'$
to ($\Gamma$) at $M$.  We will come back to this property later on.

	In order to test the validity of the continuum  approximation,
let us consider the initial  profile of Fig.\  \ref{fig1}  a such that
all  initial  terraces  have the same width  except the first one.  If
there  is a total  normal  Schwoebel  effect  such  that  there  is no
interaction   between  a  step  and  its   upper   terrace,   that  is
$\varphi_{l}(l)=0$,  then all  steps  will  move at the same  constant
velocity, and therefore terrace widths will remain constant.  In other
case, the  interaction  of the second  step with a different  one will
induce a modification  in the step  velocity, and then the velocity of
the  steps  will  change  in time  and will be  different  from one to
another.  The  continuum   approximation  is  not  sensitive  to  this
qualitatively  different  behaviour  induced by the Schwoebel  effect.
{}From equation (\ref{5}), one sees that the function $\phi(x)$ does not
change qualitatively whether  $\varphi_{l}(-x')$ is zero or not.  This
effect will be studied quantitatively in the next section.

\section{Upper ledges at long times}

	We are now concerned with the evolution of the simple  profile
already  considered  in section 2.  At the  beginning,  the  distances
$l_{1},   l_{2},...$   are   finite   and  equal.  It  will  be  shown
self-consistently  that the  distances go to infinity  for long times.
We take this as an Ansatz  to be proved  later.  For  large  values of
$l$, equations (\ref{1var1}) and (\ref{1var2}) read

\begin{eqnarray}
\varphi_{l}(l)&=&\Delta (A'-2 e^{- \kappa l}+C' e^{-2 \kappa l })
						\label{aap1} \\
\varphi_{r}(l)&=&\Delta (A''-2 e^{- \kappa l}+C'' e^{-2 \kappa l })
							\label{aap2}
\end{eqnarray}

\noindent  where  $\Delta$,  $A'$,  $A''$, $C'$ and $C''$ are given in
appendix A.  It is  necessary to go to second  order in  $\exp(-\kappa
l)$  because,  when $D''$ goes to zero, $A''$ and $C''$ go to infinity
while  $\Delta$  goes to zero.  A  similar  effect  takes  place  when
$D'=0$.

	From (\ref{1}), one obtains

\begin{equation}
\dot{l}_{n}=\dot{x}_{n+1}-\dot{x}_{n}=-\varphi_{l}(l_{n})-
\varphi_{r}(l_{n+1}) +\varphi_{l}(l_{n-1})+\varphi_{r}(l_{n})
						\label{aap3}
\end{equation}

	Combining   eq.(\ref{aap3})   with   eqs.   (\ref{aap1})   and
(\ref{aap2}), one obtains for $n>1$

\begin{equation}
\dot{l}_{n}=\Delta \left(2 e^{-\kappa l_{n+1}}-2 e^{-\kappa l_{n-1}}
- C' e^{-2 \kappa l_{n}}+
C' e^{-2 \kappa l_{n-1}}-C'' e^{-2 \kappa l_{n+1}}+C''
e^{-2 \kappa l_{n}}\right)
						\label{aap4}
\end{equation}

\noindent and for $n=1$:

\begin{equation}
\dot{l}_{1}= \Delta \left(2 e^{-\kappa l_{2}}+(C''-C')
e^{-2 \kappa l_{1}}- C'' e^{-2 \kappa l_{2}}\right)
						\label{aap5}
\end{equation}

	If $C'$ and $C''$ are finite, the terms in $\exp(-2 \kappa l)$
may be neglected in (\ref{aap5})  for large $l$.  Then these equations
turn out to have solutions of the form

\begin{equation}
l_{n}(t)=\frac{1}{\kappa} \ln (B_{n} t)		         \label{aap6}
\end{equation}

	Indeed, insertion of (\ref{aap6}) into (\ref{aap4}) yields

\begin{equation}
\frac{1}{B_{n+1}}=\frac{1}{B_{n-1}}+\frac{1}{2 \Delta \kappa}
				      \label{bes}
\end{equation}

\noindent while relations (\ref{aap6}) and (\ref{aap5}) yield

\begin{equation}
\frac{1}{B_{2}}=\frac{1}{2 \Delta \kappa}        \label{bes2}
\end{equation}

	From (\ref{bes}) and (\ref{bes2}) one deduces

\begin{eqnarray}
\frac{1}{B_{2 n}}&=&\frac{n}{2 \kappa \Delta} \label{bes3} \\
\frac{1}{B_{2 n+1}}&=&\frac{1}{B_{1}}+\frac{n}{2\kappa \Delta}
						\label{bes4}
\end{eqnarray}

\noindent  The numerical  solution  (section 5) shows that the correct
solution is the most symmetrical  one, namely $B_{1}=4 \kappa \Delta$,
so that

\begin{equation}
l_{n}(t)\simeq \frac{1}{\kappa} \ln \left( \frac{t}{n}\right)
						\label{ln22}
\end{equation}

	 Under  total   normal   Schwoebel   effect,   $D''=0$,   then
$\varphi_{l}=0$,  as seen from  (\ref{1var1}).  Then it  follows  from
(\ref{aap3})                                                      that
$\dot{l}_{n}=-\varphi_{l}(l_{n+1})+\varphi_{l}(l_{n})$   and,  if  all
$l_{n}$'s are equal at the beginning,  they remain equal, as expected.
If  total  inverse  Schwoebel  effect  is  considered,   $D'=0$,  then
$\varphi_{r}=0$ as seen from (\ref{1var2}), and for $\kappa l >>1$ and
$\kappa D<D''$:

\begin{equation}
\varphi_{l}(l)=(\rho_{0}-F\tau_{0})\tanh (\kappa l)\simeq
(\rho_{0}-F\tau_{0}) (1-2 e^{\kappa l})
                   	\label{asym2}
\end{equation}

	In this case, expression (\ref{aap3}) reduces to

\begin{equation}
\dot{l}_{n}=-\varphi_{l}(l_{n})+\varphi_{l}(l_{n-1})
						\label{asym4}
\end{equation}

\noindent which has a solution similar to eq.(\ref{ln22})

\begin{equation}
l_{n}(t)=\frac{1}{2 \kappa}\ln \left(\frac{t}{n}\right)	\label{asym5}
\end{equation}

	These logarithmic solutions correspond to a self-similar shape
of the surface, since  $l_{\alpha  n}(\alpha t)$ $ = l_{n}(t)$ for any
value of the parameter $\alpha$.

\section{Piecewise linear approximation}

\subsection{The approximation}

	In this section we are interested in the evolution of the same
profile as the one  considered  in the  previous  section, but we will
focus  our  attention  on  the  short  time  evolution.  We are  again
interested in solving eqs.(\ref{1})  subject to the initial conditions
(Fig.\ \ref{fig1} a)

\begin{eqnarray}
x_{n}-x_{n-1}&=&l(0) \;\;\;\;\;\;\;\; , n>1
						     \nonumber\\
x_{1}-x_{0}&=&\infty                                  \label{ci}
\end{eqnarray}

\noindent  with $l(0) < 2/\kappa$, so that the  saturation  regime has
not been reached.  Otherwise, the result of the previous section would
apply.

	In order  to  solve  the  evolution  equations,  we will  take
advantage of the form of the hyperbolic tangent, which is approximated
by the piecewise linear function

\begin{equation}
\varphi(\xi)=\left\{\begin{array}{ccc}\xi \;\;\;\;\;\;\;\; ,\xi\le 1\\
                     1 \;\;\;\;\;\;\;\; ,     \xi\ge 1
              \end{array}    \right.
              				        \label{aprox}
\end{equation}

	 We  will  study  the   evolution   of  the  initial   profile
eq.(\ref{ci})    within    the    piecewise    linear    approximation
eq.(\ref{aprox})  in both the symmetric  model and under total inverse
Schwoebel effect.

\subsection{Symmetric model}

	In this case, when all $l_{n}$'s are smaller than  $1/\kappa$,
using the form of our  initial  conditions  (\ref{ci}),  eqs.(\ref{1})
read

\begin{eqnarray}
\dot{x}_{1}&=&-\frac{A\kappa^{2}}{2}(x_{2}-x_{1})-A\kappa \nonumber\\
\dot{x}_{n}&=&-\frac{A\kappa^{2}}{2}(x_{n+1}-x_{n-1})\;\;\;\;\;\;\;\;
								, n>1
							  \label{neq}
\end{eqnarray}

\noindent  where  $A\equiv D (\rho_{0}-F  \tau_{0})$.  As usual, it is
convenient  to rewrite the  evolution  equations  in terms of the step
widths,

\begin{eqnarray}
\dot{l}_{1}(t)&=&-\frac{A\kappa^{2}}{2}l_{2}+A\kappa
							 \nonumber\\
\dot{l}_{n}(t)&=&\frac{A\kappa^{2}}{2}(l_{n-1}-l_{n+1})
					\;\;\;\;\;\;\;\; , n>1
						 \label{wneq}
\end{eqnarray}

	The solution of eqs.(\ref{wneq}), as derived in appendix B, is
of the form

\begin{eqnarray}
l_{2 n}(t)&=&\frac{2}{\kappa}-\left[
J_{0}(A\kappa^{2}t)+J_{2n}(A\kappa^{2}t)+2\sum_{i=1}^{n-1}
J_{2i}(A\kappa^{2}t)\right] \left( \frac{2}{\kappa} -l(0)\right)
					                \nonumber\\
l_{2n+1}(t)&=&l_{1}(A\kappa^{2}t)-\left[J_{1}(A\kappa^{2}t)+J_{2n+1}
(A\kappa^{2}t)+\right. \nonumber\\
&&\left. +\sum_{i=1}^{n-1}J_{2i+1}(A\kappa^{2}t)
\right] \left(\frac{2}{\kappa}-l(0)\right) \;\;\;\;\;\;\;\;\;\;\;\;\;
, n\geq 1						 \nonumber\\
l_{1}(t)&=&l(0)+\left[J_{1}(A\kappa^{2}t)+
2\sum_{i=1}^{\infty}J_{2i+1}(A\kappa^{2}t)\right]
\left(\frac{2}{\kappa}-l(0)\right)
						 \label{soltc}
\end{eqnarray}

	This solution shows that at the beginning the first step moves
faster  than  the  others,  and the  rest  of the  steps  move  faster
progressively.

	However, the validity of eq.(\ref{soltc}) is restricted by the
fact  that in  eq.(\ref{neq})  we have  assumed  all  $l_{n}$'s  to be
smaller than  $1/\kappa$ .  Thus, when the width of one step surpasses
the  characteristic  length  $2/\kappa$ this solution will break down.
The first step to fulfill this requirement is the first step, $l_{1}$,
for a time $t\sim  2.65/A\kappa^{2}$.  At longer times,  (\ref{soltc})
does not apply.

\subsection{Total inverse Schwoebel effect}

       The piecewise linear  approximation  will now be applied to the
case  $\varphi_{r}=0$   within  the  piecewise  linear  approximation.
Taking the initial  conditions  given by  eq.(\ref{ci}),  as $\kappa D
<<D''$, the evolution equations read, at least for small times,

\begin{eqnarray}
\dot{x}_{1}&=&-A \kappa
					             \nonumber\\
\dot{x}_{n}&=&-A\kappa^2 (x_{n}-x_{x-1}) \;\;\;\;\;\;\;\;\;\;\;\;\; ,
	n\geq 2
				            \label{left}
\end{eqnarray}

        In terms of the widths, one can rewrite these equations as

\begin{eqnarray}
\dot{l}_{1}&=&A \kappa (1-\kappa l_{1})
							 \nonumber\\
\dot{l}_{n}&=&A \kappa^2 (l_{n-1}-l_{n})\;\;\;\;\;\;\;\;\;\;\;\;\; ,
						n\geq 2
						  \label{leftw}
\end{eqnarray}

\noindent  which, using as initial  conditions  eqs.(\ref{ci}),  has a
solution of the form

\begin{equation}
l_{n}(t)=\frac{1}{\kappa}-\left[\frac{1}{\kappa}-l(0)\right]
e^{-A \kappa^2 t} \sum_{p=1}^{n-1}\frac{(A \kappa^2 t)^{p}}{p!}
						      \label{ss}
\end{equation}

	In contrast with eq.(\ref{soltc}),  this solution turns out to
be valid for all times,  because  $l_{n}(t)<1/\kappa$  for all $t$ and
$n$, and therefore  the regime of the previous  section,  valid in the
symmetric  model, is never achieved in the completely  asymmetric one.
However,  its  behaviour  is  qualitatively  similar to the  symmetric
model.  At short times the first step begins to move faster, while the
others  move at the same  speed.  After  some  time, the  second  step
increases  its  velocity,  separates  from  the  following  steps  and
approximates the first step, and so on.  This behaviour is easily seen
studying   the   difference   in   width   of  two   adjacent   steps,
$l_{n}-l_{n-1}$.  This  difference  shows a  maximum  at  time  $t=n$.
However, the height of the peak decreases with  increasing  $n$, which
means that this effect decreases quantitatively with $n$.

\section{Numerical results}

	The nonlinear character of the evolution of the crystal shape,
as  shown  by  eq.(\ref{1})  makes  it  impossible  to find a  general
analytic solution of the shape as a function of time.  In the previous
sections  we  have  found  such  analytic   expressions  in  different
asymptotic  limits.  However, in order to study the  evolution  of the
profile at any time, it is necessary to carry out a numerical study of
the  system  (\ref{1}).  The  results  will  enable  us to  check  the
different approximations  introduced so far, both in the continuum and
in the discrete cases.

	We have solved the  differential  set of  equations  (\ref{1})
corresponding to the discrete model in the symmetric case, and initial
conditions  given  by  eq.(\ref{ci}).  We  have  used a  fourth  order
Runge-Kutta  algorithm  \cite{fla}.  We have  considered  a set of 200
steps.  The   neighbouring   left-side  step  to  our  first  step  is
considered  to be at infinity.  The rightest one is assumed to move at
constant  speed, which is our  boundary  condition  and means that the
results  are  obtained  in a  reference  system  which  moves with the
initial slope of the steps.  This  boundary  condition  means that the
terraces at the right of the last one have not  changed  significantly
their  width.  This will be true  only for a certain  time  scale.  As
soon as the  width of the  rightest  terrace  begins  to  evolve,  our
solution  will  be  wrong.  From  the  analytic  results  of  previous
sections, it may be argued that at $t\neq 0$ all terraces change their
width.  However,  this change is extremely  small and, as explained in
the previous subsection, only the first ones change  significantly, so
it takes some time until the terraces at the bottom change their width
appreciably.  This fact gives us an easy way to check the  validity of
the  solution.  It turns out that all relevant  features, at least for
the first  terraces,  take  place  during the time  scale in which the
numerical  solutions  subject  to our  boundary  condition  are valid.
However,  the  numerical  study at very short times is obscured by the
discrete time step.

	The differential  equations are solved for different values of
the initial  width, $l(0)$, as well as different time steps,  checking
the stability of the solution against computational artifacts.

	 Figure\ \ref{fig3} shows the profile at different times.  Due
to the fast  increase of the width of the first steps with  respect to
the other steps, it is necessary to choose two different length scales
in the vertical and horizontal axes.  In the vertical axis we take the
step  height  as  the  unit  length  while  in  the  horizontal   axis
$\kappa^{-1}$  is considered  the unit length.  This is the reason why
the  initial  profile  appears as an almost  vertical  straight  line.
Moreover, $2 A \kappa^{2}$ is the unit time  throughout  the numerical
calculations.  The  curves  are  plotted  every  9 time  steps,  which
corresponds to a plot every $\Delta t=0.01$.

	We  have  also  studied  the   behaviour  of  the  surface  at
intermediate times.  In this regime the numerical solution agrees with
the expressions (\ref{aap6}) in the symmetric model, and (\ref{asym5})
in the total inverse Schwoebel model, supporting the symmetry argument
we have  employed  in order  to  derive  such  equations. In  Figure\
\ref{fig4} we show the width of the first three terraces as a function
of time, and both the logarithmic dependence and the numerical factors
are  recovered  exactly  after a short  transient in both models.  The
difference in the slope is easily  appreciated.This good matching with
the analytic  predictions is again in agreement with the idea that the
solution  (\ref{soltc})  is  valid  for  short  times,  but  that  the
saturation  regime  is  achieved  in few  time  units  (in  the  above
mentioned units).

	We can  compare  the  shape of the  discrete  model  with  the
predictions  of the continuum  approximation.  In Fig.\  \ref{fig3} we
have  checked the validity of Frank's  construction  for the  discrete
model.  We find that all lines joining  points of the crystal  surface
where the slope has a given  value at  different  time meet at a point
$O$,  which  turns  out to be the  angular  point at  $t=0$.  Thus all
shapes are homothetic of a particular one, and the homothety center is
$O$.

	 Finally, in Figure\  \ref{fig5}  the  self-similarity  of the
profile is  checked.  Neglecting  a short  transient  one  observes  a
perfect  scaling of the whole  profile.  The continuum  model predicts
three scaling regions, corresponding to the two initial straight lines
and the part of the  profile  that due to the  dynamics  has  deviated
significantly from the other two.  Therefore, two different crossovers
are  expected.  However,  they are not  readily  observed  numerically
because of the finite time step which produces a spurious  propagation
front which blurres such a crossover.

\section{Grooves}

In this section we will consider the evaporation of a solid limited by
a periodic  array of grooves  (Fig.\  \ref{fig6} a).  The sides of the
grooves will be assumed to be planar at the initial time $t=0$.  It is
sufficient to study the evolution of a half period  (Fig.\ \ref{fig6}
b), which contains a time-dependent number $n_{max}(t)$ of steps.  The
origin $x=0$ will be chosen at the highest  point,  supposed to be the
left hand side of the  half-period.  The  steps  will be  labelled  1,
2,...,$n_{max}$ from the left to the right.

	Only  the  symmetric  case  will be  considered,  so that  the
evolution of an isolated step is governed by equations  (\ref{1})  and
(\ref{2}).  The two  lowest  steps of each  period  deserve  a special
attention  because they move in opposite  directions,  so that a facet
appears at the lowest  parts of the profile  (Fig.  \  \ref{fig6}  c),
Therefore, the position $n_{max}$ of the last step depends on the time
t.

	It may be of interest to rewrite the equations of motion for
the  uppest and lowest steps as:

\begin{eqnarray}
\dot{x}_{1}&=&-A\kappa \left[\tanh\left(\kappa \frac{x_{2}-x_{1}}{2}
\right)+ \tanh(\kappa x_{1})  \right]
                                                \label{pyr1}\\
\dot{x}_{n_{max}}&=&-A\kappa \left[\tanh\left(\kappa (\frac{L}{2}+
x_{n_{max}})\right)+
\tanh\left(\kappa\frac{x_{n_{max}}-x_{n_{max}-1}}{2}\right)\right]
\label{pyr2}
						\end{eqnarray}

\noindent  where  $L$ is the  initial  width  of the  groove.  Two new
features  appear with respect to the evolution of the profile  studied
in  the  previous  sections.  On  one  hand,  the  highest  step  will
disappear  after some time, and at that time it is necessary to update
the  $x_{n}$'s  and to replace  $n_{max}$  by  $n_{max}-1$.  A similar
updating will take place every time $x_{1}$  vanishes,  until the last
step  disappears  and the surface  becomes  perfectly  smooth.  On the
other hand, since a facet appears at the bottom, the first term at the
right hand side of equation  (\ref{pyr2})  becomes  large.  Therefore,
the lowest  step moves  faster  than the other  steps and, after  some
time, it reaches the step just above it.  Evaporation  generates  step
bunches!  Step  bunching  is  frequently  observed  after  growing  or
annealing crystals, and various  explanations have been proposed.  The
present one, though very simple, seems to be new.

	When the  lowest  step  has  reached  the next  one,  equation
(\ref{pyr2})  should be modified,  because the step  positions  should
satisfy the conditions $x_{n}>x_{n-1}$.  The new equation may be found
if one assumes, following BCF, that steps (forming bunches or not) are
in equilibrium  with the bulk.  This results from the assumption  that
the motion of atoms along steps  requires  lower  activation  energies
than  atom  detachment  from  steps.  It  follows  that  the  chemical
potential on the lowest  terrace (the broad one) near a step should be
the bulk chemical potential, and that the adatom density on the lowest
terrace near a step should be the equilibrium  adatom density, just as
in the case of a single  step.  On the other hand, the adatom  density
$\rho(x)$  satisfies  the  diffusion  equation  (A.1)  (see  appendix)
independently of the number $p$ of steps in the bunch which limits the
lowest terrace.  Since the boundary  condition is also  independent of
$p$,  $\rho(x)$ is  independent  of $p$.  Therefore the current at the
right hand side of the lowest  step, which is the  gradient of $\rho$,
is    also    independent    of    $p$,    and    therefore    it   is
$\tanh(\kappa(L/2-x_{n_{max}}))$.  On the other  hand, the  current to
the  left of the  highest  step in the  lower  bunch is  $\tanh(\kappa
(x_{n_{max}-p}-x_{n_{max}-p+1})/2)$  independently of $p$.  There is a
current of atoms  inside the bunch, the effect of which is to maintain
the chemical  potential uniform within the bunch.  It is seen that the
current  from the lowest  step is larger  than from the  highest  one.
This  ensures the  stability of the bunch, i.e., its highest step does
not move faster than the lowest one.  Therefore all velocities  within
the bunch are the same:

\begin{equation}
\dot{x}_{n_{max}}=-\frac{A\kappa}{p}\left[
\tanh\left(\kappa(\frac{L}{2}- x_{n_{max}})\right)+\tanh\left
(\kappa\frac{x_{n_{max}-p}-x_{n_{max}-p+1}}{2}\right)\right]
\label{pyr3}
\end{equation}

	If bunches of steps  appear at other places than the bottom of
the profile, equation  (\ref{1}) should be modified and replaced by an
equation similar to (\ref{pyr3}).  However, no large bunches have been
observed  in the  numerical  solution  of the  equations,  so that the
resulting  profile is essentially that represented on Fig.\ \ref{fig6}
d.  If a bunch  appears, it is necessary  to check its  stability.  As
said above, a bunch is stable if the  current  from its lowest step is
larger than the current  from its highest  step.  Thus, the  condition
for a bunch of $p$ steps to be stable if its highest  step is step $m$
is

\begin{equation}
\tanh\left(\kappa \frac{x_{m}-x_{m-1}}{2}\right)<\frac{1}{p-1}
\tanh\left(\kappa \frac{x_{m+p+1}-
x_{m+p}}{2}\right)
\label{pyr4}
\end{equation}

	We have  performed a numerical  study of the evolution for the
profile shown in Fig.\ \ref{6} a, with $l(0)=0.1  \kappa$.  Initially,
the  highest  and lowest  terraces  are twice as broad as the  others.
Fig.\  \ref{fig7}  depicts  the  profile  at  different  times  during
evaporation.  The top of the  profile  remains  linear  while,  at the
bottom, a bunch  appears as  expected.  At the  beginning,  this bunch
grows, but, in our  calculations, it reaches a maximal size after some
time.  This  size is  reached  when the  width of the  lowest  terrace
becomes of order  $1/\kappa$  so that  further  increase of that width
does not produce a further  increase  of the  current  from the lowest
step.  Thus, the velocity of the bunch is of order $1/p$  according to
(\ref{pyr3}).  This  velocity  should be about the same as that of the
next step, which is of order l.  Therefore, the number $p$ of steps in
the  bunch is given  by  $p=1/(l(0)  \kappa)$.  After  the  bunch  has
reached its maximal  size, a pairing of steps is observed.  We explain
the  formation  of a step pair just above the bottom bunch as follows.
Just after  reaching  saturation,  the bunch can have  (because of the
high $p$ value in equation  (\ref{pyr3})) a velocity  smaller than the
next step.  Therefore  this step will move at a higher  velocity  than
the  others  and will  form a pair of steps  with the  next  one.  The
formation of the other step pairs are  presumably  of similar  origin.
Step  pairing  is not  always  observed  and  depends  on the  initial
conditions.

	 It is easily  deduced from the above  arguments  that, if the
initial width is larger than $1/\kappa$ no bunches are observed.

	It is interesting to compare our results (Fig.\ \ref{fig6} and
\  \ref{fig7})   with  what  might  be  expected  from  the  continuum
approximation.  For the sake of simplicity and because it  corresponds
to usual  experiments, only the case  $l<<1/\kappa$ will be addressed.

	Then, equation (\ref{6}) reduces to

\begin{equation}
\dot{z}=A\kappa^2					\label{pyr5}
\end{equation}

\noindent so that  evaporation just translates the profile  downwards.
However, the lowest  terrace does not evaporate, so that the resulting
profile  is  essentially   that  of  Fig.\  \ref{fig6}  c.  A  similar
conclusion would be obtained from Frank's construction.  The continuum
approximation  fails to  predict  the  bunches  observed  in  Figures\
\ref{fig6}  d  and\  \ref{fig7}.  As in  the  previous  sections,  the
failure of the continuum  approximation is related to the existence of
angular  points,  namely those at the bottom of Figure\  \ref{fig6} c.
However, in the profile  studied in the previous  sections the initial
angular  point  became  smooth,  and  therefore  Frank's  construction
applied  at  latter  times,  while in the  sawtooth  profile  of Fig.\
\ref{fig6},  the  angular  points  remain and even give rise to facets
which are unexpected in the continuum approximation.

	We have also  studied the  evolution  of initially  sinusoidal
profiles or other  differentiable  profiles  (Fig.\  \ref{fig8} a and\
\ref{fig8}  b).  The main new feature is that an angular point appears
at the top if the upper  terrace  is broader  than  $1/\kappa$  (Fig.\
\ref{fig8} a).  The reason is the saturation  effect  associated  with
the hyperbolic  tangent in (\ref{2}):  the top of the profile does not
evaporate  while  the  remainder  of  the  surface  follows   equation
(\ref{pyr5}).  This angular point will nevertheless not be observed in
most of real materials because large surface adlacunes (not taken into
account in the present model) will be nucleated.

	We have also made  preliminary  studies  of the  evolution  of
periodic  profiles  with  Schwoebel  effect.  The  situation is rather
complicated  because of  instabilities,  reminiscent of those found by
Kandel and Weeks \cite{Wee} in a model combining an extreme  Schwoebel
effect, growth and impurities.

\section{Conclusion}

	We have  studied  the  evaporation  of a  crystal  within  the
Burton-Cabrera-Frank  model  in two  cases:  i) when  the  surface  is
limited by two planes, one of them having a high-symmetry  orientation
(Fig.\ \ref{fig1} a); ii) for a "grooved" surface (Fig.\  \ref{fig6}).
We have paid  attention  to the  Schwoebel  effect  (asymmetry  of the
sticking coefficient) and to the validity of Frank's theorem, based on
the continuum  approximation.  However, we have not investigated cases
where the  instabilities  described  by  Schwoebel  \cite{Sch}  and by
Kandel and Weeks \cite{Wee} occur.

	In  the  case  of  a  corner  (Fig.\  \ref{fig1}  a),  Frank's
construction  can only predict the  evolution  of the crystal  above a
particular  plane  $\Pi$,  and below  another  plane  $\Pi'$,  but not
between the two planes.  Our  numerical  solution  shows a rounding of
the corner and an  evolution  of the  profile  towards a  self-similar
shape.  We have also  shown that the  distance  between  upper  ledges
diverges  logarithmically  with  time  except  in the  case of a total
normal  Schwoebel  effect.  This  divergence  wouldn't  occur  if  the
continuum  approximation  were  exact.  An  approximate  form  of  the
equation of motion has also been  investigated,  which allows an exact
solution.

	In the case of a grooved surface, the bottom of the grooves is
found to flatten, but their edges become steeper due to step bunching.
The  mechanism  responsible  for this  effect  is a very  simple  one,
however it cannot be deduced from the continuum approximation.

	The  present  work can help to  understand  the  formation  of
defects  during   annealing  of  crystal   surfaces.  However,  direct
comparison  with  experiment  is not possible  because, in the present
work,  steps  are  assumed  to  be  straight,  and  dislocations   are
neglected.  This is  only  correct  for  small  lengthscales,  usually
around 0.01 cm or less.  We have also ignored the effect of vacancies,
which are  certainly  important  near the  melting  point of  elements
\cite{Tro}, and can even suppress the saturation  effect  appearing in
formula (\ref{2}) when $\kappa l>1$.  Note that this saturation effect
has been observed experimentally in some materials, as for example did
Keller in NaCl  \cite{Kel}.  Neglecting  vacancies is only  correct if
terrace sizes are small.A  quantitative  discussion  has been given in
Ref.  \cite{Pim}.

	It is  interesting  to  compare  our  results  with  the  ones
reported  by  Stoyanov  \cite{Sto}  for a stepped  surface in which an
external force acts on the adatoms.  In that case, bunch  formation is
predicted  when   $\Psi(l)\equiv   \varphi_{r}(l)-\varphi_{l}(l)$   is
positive for a terrace wider than its neighbouring  terraces, of width
$l$.  The sign of $\Psi(l)$  depends on the  direction of the external
force.  In our case, for an initial surface given by Fig.\  \ref{fig1}
a,  $\Psi(l)$  is positive  in the case of inverse  Schwoebel  effect.
Bunches have not been  observed  because  they will develop  above the
terrace  which is  larger  than its  neighbouring  ones.  In our case,
these would correspond to terraces on the top of our first terrace and
have  not  been   considered  in  the  analysis.  What  regards  bunch
formation  in grooves,  it is  observed  even in the  symmetric  case,
although  $\varphi_{r}(l)=\varphi_{l}(l)$, because $\Psi(l)$ is always
positive for the terrace at the bottom.

\acknowledgements

	One  of the  authors  (I.P.)  wants  to  thank  Ministerio  de
Educaci\'{o}n y Ciencia for financial support.

\appendix

\section{Velocity of the Steps with Schwoebel Effect}

	In this  appendix  we  derive  the  expression  for  the  step
velocity  as  sketched  in section  1.  If the  diffusion  process  of
adatoms  on the  surface is much  faster  than the motion of steps, as
supposed  by BCF  \cite{BCF},  the adatoms  reach a  stationary  state
during the motion of the steps and the velocity of the step is related
to the flux of such an adatom density.  Then, we should  determine the
density  profile  on the  terraces  in  order to  calculate  the  step
velocity.

Between  two steps, the adatom  density  $\rho_{n}(x)$  satisfies  the
following equation

\begin{eqnarray}
\frac{\partial \rho_{n}(x)}{\partial t}&=&D \frac{\partial ^{2}
\rho_{n}(x)}{\partial x^{2}} -\frac{1}{\tau_{0}}
\rho_{n}(x) + F                \nonumber\\
&& \hspace{8.5cm}   x_{n}<x<x_{n+1}
                                                          \label{a1}
\end{eqnarray}

\noindent  where in the right hand side the first term  corresponds to
diffusion,  the  second  one to  evaporation  and  the  third  one  to
deposition.  In the present article, $F$ is taken to be zero.

	The boundary  conditions at the steps are the following  ones,
which state the  equality of two  expressions  of the current  density
$j(x)$:

\begin{eqnarray}
-j(x)&=&D\frac{\partial \rho_{n}(x)}{\partial
x}=D''\rho_{0}-D''\rho_{n}(x) \;\;\;\;\; (x=x_{n}-\epsilon)
					\label{a2} \\
-j(x)&=&D\frac{\partial \rho_{n}(x)}{\partial
x}=D'\rho_{0}-D'\rho_{n}(x) \;\;\;\;\; (x=x_{n-1}+\epsilon)
					\label{a3} \\
\end{eqnarray}

\noindent where $\epsilon$ denotes a small quantity.  $D'\rho_{0}$ and
$D''\rho_{0}$  are the current of adatoms detaching from both sides of
the step.  These values are imposed by detailed balance.

	Diffusion is usually much faster than step motion, so that the
left  hand side of  (\ref{a1})  may be  replaced  by zero.  It is then
straightforward to calculate  $\rho_{n}(x)$, and to deduce the current
density  on  both  sides  of  each  step  using   eqs.(\ref{a2})   and
(\ref{a3}),  which  are  precisely  $\varphi_{l}$  and  $\varphi_{r}$.
Equations (\ref{1var1}) and (\ref{1var2}) are therefore easily deduced
after some algebra.

When $l$ is large, the second  order  expansion of  (\ref{1var1})  and
(\ref{1var2})   in   powers   of   $\exp(-\kappa   l)$  is  given   by
eqs.(\ref{aap1}) and (\ref{aap2}) where

\begin{eqnarray}
\Delta&=&\frac{(\rho_{0}-F\tau_{0}) \kappa D}{1+ \frac{\kappa^{2}
D^{2}}{D'D''} +\frac{\kappa D}{D'}+\frac{\kappa D}{D''}}
					\label{a4} \\
A'&=&1+\frac{\kappa D}{D'}		\label{a5}\\
C'&=&1-\frac{\kappa D}{D'}-\left(1+\frac{\kappa D}{D'}\right)
\frac{\frac{\kappa D}{D'}+\frac{\kappa D}{D''}-1-
\frac{\kappa^{2}D^{2}}{D'D''}}{1+ \frac{\kappa^{2}
D^{2}}{D'D''}+\frac{\kappa D}{D'}+\frac{\kappa D}{D''}}
					\label{a7}
\end{eqnarray}

\noindent and $A''$ and $C''$ are obtained by  interchanging  $D'$ and
$D''$ in (\ref{a5}) and (\ref{a7}) respectively.

\section{Short-time Symmetric Case Profile}

	In this appendix we derive  eqs.(\ref{soltc}),  starting  from
the approximate terrace width evolution equations (\ref{wneq}).  Using
vector notation these are rewritten in the more convenient form

\begin{equation}
|\dot{l}(t)> = \frac{1}{2} A \kappa^{2} Y |l(t)> +
A \kappa |1>
                                                        \label{b9}
\end{equation}

\noindent with the vector $|\dot{l}(t)>$ being

\begin{equation}
|\dot{l}(t)>=\left( \begin{array}{l}
                     l_{1}(t)\\
                     l_{2}(t)\\
                     \vdots\\
                     l_{n}(t)\\
                     \vdots\\
                     l_{N}(t)
                    \end{array}       \right)          \label{bb}
\end{equation}

\noindent and where we have introduced

\begin{equation}
Y=\left( \begin{array}{lrrrrr}
           0 & -1 & 0 & 0 & 0 & ...\\
           1 & 0 & -1 & 0 & 0 & ...\\
           0 & 1 & 0 & -1 & 0 & ...\\
           0 & 0 & 1 & 0 & -1 & ...\\
           0 & 0 & 0 & 1 & 0 & ...\\
           ...&...&...&...&...&...
           \end{array}            \right)
           \hspace{1cm}\mbox{and}\hspace{1cm} |1>=\left(
                                  \begin{array}{c}
                                       1 \\0 \\0 \\0 \\0 \\ ...
                                      \end{array}          \right)
                                                       \label{b10}
\end{equation}

        The formal solution of (\ref{b9}) is easily checked to be

\begin{equation}
|l(t)>=e^{\Omega t}\left( |l(0)>+2\kappa^{-1}Y^{-1} |1>\right)
-\frac{2}{\kappa}Y^{-1} |1>
                                                         \label{b11}
\end{equation}

\noindent with $\Omega=\frac{1}{2} A \kappa^{2}Y$.

	In order to find an explicit  solution to  eq.(\ref{b11}),  we
should   decompose  the  matrix  $Y$  in  eigenvectors,  so  that  the
exponential  $e^{\Omega  t}$ can be  diagonalized.  To  this  end,  we
introduce the vectors

\begin{equation}
<k|=\left(\begin{array}{ccccccc}
        1, & e^{-i k}, & e^{-2 i k}, & \ldots, & e^{-i n k}, & \ldots,
          & e^{-i N k}
          \end{array}                    \right)
                                                          \label{b12}
\end{equation}

\noindent  with $n$ a  natural  number,  $n=0,1,\ldots,N$.  We can now
construct the eigenvectors of $Y$, which is a finite Toeplitz  matrix,
as an appropriate combination of $|k>$.  It is readily checked that

\begin{equation}
|k)\equiv \frac{1}{\sqrt{2 N}} \left( e^{i k} |k>+e^{-i k}
|\pi-k>\right)
                                                          \label{b13}
\end{equation}

\noindent  are such vectors if $N$ is an odd number, and $k$ takes the
values   $k=\frac{(2   j-1)   \pi}{2   (N+1)},\;   j=-N,...,N+1$.  The
corresponding eigenvalues are $-2 i \sin (k)$.  Then, the exponential,
in terms of these eigenvectors, has the form

\begin{equation}
e^{\frac{1}{2}A \kappa^{2} t Y}= \sum_{k} e^{-i A\kappa^{2} t
\sin (k)} |k) (k|
                                                         \label{b14}
\end{equation}

	We should now express the  exponential in real space.  To this
end, we have to evaluate the action of the operator on vectors  $|n>$,
which has its n-th component equal to 1, and the rest of them equal to
zero.  Using the fact that

\begin{equation}
<n|k)=\frac{1}{\sqrt{2 N}}\left( e^{i n k}+ (-1)^{n-1} e^{-i n k}
\right)				\label{b15}
\end{equation}

\noindent  one can  determine the value of the  exponential  acting on
such vectors

\begin{eqnarray}
&&<n|e^{\frac{1}{2}A \kappa^{2} t Y}|m>=\sum_{k}
e^{-i A\kappa^{2} t\sin (k)} <n|k)(k|m>
                                                 \nonumber\\
&&=\frac{1}{2 N} \sum_{k} e^{-i A\kappa^{2} t\sin (k)} \left( e^{-i n
k}+(-1)^{n-1} e^{i n k}\right) \left( e^{i m k}+ (-1)^{m-1}
e^{-i m k} \right)
                                                        \label{b16}
\end{eqnarray}

	As we are  interested in the situation  where there is a large
number of terraces,  and then the  behaviour of the system will not be
very  sensitive  on the  specific  value of N, which is large, one can
approximate the sums in eq.(\ref{b16}) by integrals

\begin{eqnarray}
&&<n|e^{\frac{1}{2}A \kappa^{2} t Y}|m>=\frac{1}{2\pi}
\int_{-\pi/2}^{\pi/2} dk e^{-i A\kappa^{2} t\sin (k)}  \times
                                 \nonumber\\
&& \left( e^{i (m-n) k}+ (-1)^{(n+m)} e^{i (n-m) k}+(-1)^{(m+1)}
e^{-i (m+n) k}+(-1)^{(n+1)} e^{i (m+n) k} \right)
                                                        \label{b17}
\end{eqnarray}

        Then, using the equality
\begin{equation}
\frac{1}{2 \pi} \int _{-\pi/2}^{\pi/2} dk e^{-i A\kappa^{2} t\sin(k)}
(e^{i p k}+(-1)^{p} e^{-i p k})= J_{p}(-A \kappa^{2}t)
							 \label{b18}
\end{equation}

\noindent with  $J_{n}(x)$  being the Bessel  function of order n, one
determines the value of eq.(\ref{b17}), namely

\begin{equation}
<n|\exp(A \kappa^{2} t Y/2)|m>=J_{n-m}(A \kappa^{2} t)+(-1)^{m+1}
J_{n+m}(A\kappa^{2} t)  \label{b19}
\end{equation}

\noindent  which  enables  us to  write  down  the  expression  of the
exponential in real space

\begin{equation}
\exp(A \kappa^{2} t Y/2)=\left(\begin{array}{clllll}
           J_{0}+J_{2}&-J_{1}-J_{3}&J_{2}+J_{4}&-J_{3}-J_{5}&...&...\\
           J_{1}+J_{3}&J_{0}-J_{4}&-J_{1}+J_{5}&J_{2}-J_{6}&...&...\\
           J_{2}+J_{4}&J_{1}-J_{5}&J_{0}+J_{6}&-J_{1}-J_{7}&...&...\\
           J_{3}+J_{5}&J_{2}-J_{6}&J_{1}+J_{7}&J_{0}-J_{8}&...&...\\
           ...&...&...&...&...
                        \end{array} \right)
                                                        \label{b20}
\end{equation}

\noindent where all the Bessel  functions have an argument equal to $A
\kappa^{2} t$.

	On the other hand, when  applying  $Y$ on $|l(t)>$  the formal
solution (\ref{b11}) is expressed

\begin{equation}
\left(\begin{array}{c}
                -l_{2}(t)\\
                l_{1}(t)-l_{3}(t)\\
                l_{2}(t)-l_{4}(t)\\
                ...
                \end{array} \right)= \exp(A \kappa^{2} t Y/2)
                \left(\left(\begin{array}{c}
                -l(0)\\0\\0\\...\end{array}\right)+
                2\kappa^{-1}|\psi>\right)-2 \kappa^{-1}|\psi>
                                                        \label{b21}
\end{equation}

Now, applying expression  (\ref{b20}) for the matrix to eq.(\ref{b21})
we finally arrive at expressions  (\ref{soltc}) for the terrace widths
as a function of time and the number of the  terrace,once  the initial
terrace width is known.

\begin{figure}
\caption  {a)Profile of a crystal  surface made of two
semi-infinite  half planes.  The left-side  surface  corresponds  to a
high  symmetry  plane.  b)  Scheme  of  the  movement  of  adatoms  on
terraces,  showing the probability of each event.  Schwoebel effect is
taken into account through the coefficients $D'$ and $D''$.}
\label{fig1}
\end{figure}

\begin{figure}
\caption {Frank's  construction.  Left hand part:cross
section of the initial  crystal  (dashed  curve) and of the actual one
(full curve); right hand part:  the surface generated by the point $M$
defined by $\vec{\Omega M}=\vec{n}/v(\vec{n})$ .  The point $P$ of the
crystal  surface  where the normal is  $\vec{n}$  moves on the  dotted
straight line, parallel to the normal to $\Gamma$ at $M$.  On the left
hand of the dotted  line, the  crystal is planar and  parallel  to its
original  orientation.  Only the  useful  part of the curve of $M$ has
been shown.  Note the presence of a point  $M_{\infty}$  at  infinity.
It  corresponds  to  $l\rightarrow\infty$  and its  abcissa  is  $-1$,
corresponding to the formula $\vec{\Omega M}=-(1,l)/\varphi(l)$, where
$\varphi$ is given by (5).}
\label{fig2}
\end{figure}

\begin{figure}
\caption  {Snapshots  of the  evolution of the initial
profile shown in Fig.  1 a in the symmetric  case.  Profiles are drawn
every 20 times in the units defined in the text.  Two different scales
have been chosen in the axes, as  explained  in the text.  The profile
in the  discrete  symmetric  case is  shown  to be in  agreement  with
Frank's construction.  Two equal slopes at different times are seen to
be joined by a straight line, and different straight lines converge at
0, the angular point at t=0.}
\label{fig3}
\end{figure}

\begin{figure}  \caption{  Width  of the  three  first  terraces  as a
function   of  time  at  long  times.  $\triangle$,   $\diamond$   and
$\bigtriangledown$  correspond to the symmetric  case, while  $\circ$,
$\bullet$ and $\Box$ correspond to the total inverse Schwoebel effect.
The logarithmic behavior, the prefactor and the constant are recovered
in  both   cases.  The  1/2   difference   in  the  slope  is   easily
appreciated.}
\label{fig4}
\end{figure}

\begin{figure}  \caption  {Self-similarity  of  the  profiles  of  the
crystal  without   Schwoebel   effect.  The  crossovers   between  the
different regimes are blurred by numerical artifacts.}
\label{fig5}
\end{figure}

\begin{figure}
\caption {a) Initial profile of the surface studied in
Section 6.  b) Enlargement of the half-period  inside the rectangle of
Fig.  6 a.  c) The  evaporation  shape as would be predicted  from the
continuum  approximation.  d)  The  real  shape  resulting  from  step
bunching,  not  taking  into  account  the step  pairing  occasionally
observed in simulations.}
\label{fig6}
\end{figure}

\begin{figure}  \caption{  Snapshots of the  evolution  of the initial
profile  shown in Fig.1a.  Profiles  are drawn  every ten times in the
units defined in the text.  Two  different  scales have been chosen in
the axes, as explained in the text.}
\label{fig7}
\end{figure}

\begin{figure}
\caption{ Snapshots of the evolution of an initial
parabolic profile. a)  Evolution when initially all
terraces are smaller than $1/\kappa$.  b) Evolution when initially the
top terraces are larger than $1/\kappa$ and the bottom ones smaller
than this value.}
\label{fig8}
\end{figure}
\end{document}